\documentclass[prl, aps,
twocolumn,
 showpacs,amsmath,amssymb]{revtex4}
\usepackage{graphicx}
\usepackage[cp1251]{inputenc}
\usepackage{dcolumn}
\usepackage{bm}
\usepackage{amssymb}
\usepackage{amsmath}
\usepackage{verbatim}
\usepackage{color}
\usepackage{ulem}
\def \be {\begin{equation}}
\def \ee {\end{equation}}
\def \bea {\begin{align}}
\def \eea {\end{align}}
\def \p {\partial}
\def \BEA {\begin{eqnarray}}
\def \EEA {\end{eqnarray}}

\def \BC {\begin{cases}}
\def \EC {\end{cases}}

\begin{document}

\title{
Anomalous Hooke's law in disordered graphene
}
\author{I.V.~Gornyi$^{1,2,3,4}$}
\author{ V.Yu.~Kachorovskii$^{1,2,3,4}$}
\author{A.D.~Mirlin$^{1,3,4,5}$}
\affiliation{$^{1}$Institut f\"ur Nanotechnologie,  Karlsruhe Institute of Technology,
76021 Karlsruhe, Germany
\\
$^{2}$    A.F.~Ioffe Physico-Technical Institute,
194021 St.~Petersburg, Russia
\\$^3$ \mbox{Institut f\"ur Theorie der kondensierten Materie,  Karlsruhe Institute of
Technology, 76128 Karlsruhe, Germany}
\\ $^{4}$
L.D.~Landau Institute for Theoretical Physics, Kosygina
  street 2, 119334 Moscow, Russia
  \\$^{5}$ Petersburg Nuclear Physics Institute, 188300, St.Petersburg, Russia
}

\date{\today}

\begin{abstract}
The discovery of graphene,  a single monolayer of graphite, has  closed  the discussion on stability of 2D  crystals.
Although thermal fluctuations of  such  crystals  tend to destroy the long-range order in the system, the crystal
can be stabilized by strong anharmonicity effects.  This competition
is the central issue of the crumpling transition, i.e., a transition between flat and crumpled phases.
We show that
anharmonicity-controlled
 fluctuations of a graphene membrane around  equilibrium  flat phase lead to unusual elastic properties.
In particular, we demonstrate that stretching $\xi$ of a flake of graphene
is a nonlinear function of the applied tension at small tension:
${\xi\propto\sigma^{\eta/(2-\eta)}}$ and ${\xi\propto\sigma^{\eta/(8-\eta)}}$ for clean and
strongly disordered graphene, respectively. Conventional linear Hooke's law, ${\xi\propto\sigma}$
is realized at sufficiently  large tensions: ${\sigma\gg\sigma_*},$ where $\sigma_*$ depends both on
temperature and on the  disorder strength.
    \end{abstract}
\maketitle

Hooke's law (HL)---introduced  by  Robert Hooke about 350 years ago---states
that the force needed to extend or compress an elastic body
by some distance  is proportional to that distance.
Conventional theory of elasticity predicts that this law is  fulfilled  for
low fields (in the so-called ``elastic range'' of tensions) and gets violated at sufficiently large tensions.

The goal of this Letter is to explore stretching of graphene, a famous two-dimensional (2D) material
\cite{Geim,Geim1,Kim,geim07,graphene-review,review-DasSarma,review-Kotov,book-Katsnelson,book-Wolf,book-Roche},
as a reaction on applied tension.  Measurement of the elasticity of free-standing graphene is accessible to current experimental techniques \cite{lee, metten,blees15,lopez-polin15,nicholl15}.
Remarkably, we find that, for graphene,  HL fails even in the limit of the  infinitesimally small tension.
The underlying physics has a very close relation to the well known problem of thermodynamic stability of 2D crystals \cite{mermin,landau}.

Free-standing graphene is a remarkable example of an elastic crystalline 2D membrane
with a high bending rigidity ${\varkappa\simeq 1}$~eV.
The most important feature distinguishing such a membrane from conventional 2D semiconductor systems is the
existence of specific type of out-of-plane phonon modes---flexural phonons (FP) \cite{Nelson}.

In contrast to in-plane
phonons with the linear dispersion,
the FP are very soft, $\omega_{ \mathbf{q}} \propto q^2$, and, consequently,
the out-of-plane thermal fluctuations are unusually strong and tend to destroy graphene membrane by
driving it into the
crumpled phase \cite{Nelson}.
The competing  effect is
the anharmonicity
that plays here a
key role.

This question was intensively discussed more than two decades ago
\cite{Nelson,Nelson0,Crump1,NelsonCrumpling,david1,buck,Aronovitz89,david2,lower-cr-D2,d-large,disorders,disorder-imp,Gompper91,
RLD,Doussal,disorders-Morse-Grest,RLD1,Bowick96}
in connection
with   biological membranes, polymerized layers, and inorganic surfaces.
The interest to this topic has been renewed more recently~\cite{eta1,Gazit1,Gazit2,Hasselmann,kats1,kats2,Amorim,kats3}
after discovery of graphene.
It was found \cite{Nelson0,Crump1,NelsonCrumpling,david1,Aronovitz89,buck,david2} that the anharmonic coupling of
in-plane and out-of-plane phonons stabilizes the membrane for not too high temperatures $T$. This is connected with a
strong
 renormalization of the bending rigidity \cite{Aronovitz89,lower-cr-D2,Doussal},
${\varkappa\to\varkappa_q\propto{q}^{-\eta},~~\text{for}~~q \to 0},
$
with a certain critical index $\eta$.
Due to the high bare value of $\varkappa$, {\it clean} graphene remains flat up to all realistic  temperatures.
The critical exponent $\eta$ was determined within several
approximate analytical schemes \cite{david1,david2,Aronovitz89,Doussal,eta1}.
Numerical simulations for a 2D membrane embedded in 3D space yield $\eta=0.60 \pm 0.10$ \cite{Gompper91}
and $\eta=0.72 \pm 0.04$ \cite{Bowick96}.

In a recent paper~\cite{my-crump}, we have developed a theory of rippling and crumpling
in {\it disordered} free-standing graphene.
We have shown that random fluctuations of the membrane curvature caused by static disorder
may strongly affect properties of the membrane. We have derived coupled renormalization-group (RG)
equations describing the combined flow of $\varkappa$ and disorder strength $b$,
determined the phase diagram (flat vs. crumpled) in the $(\varkappa, b)$ plane,
and explored the rippling in the flat phase.

In the present Letter, we explore the fate of HL in clean and disordered graphene.
We find that linear HL  breaks down both for clean and disordered cases, so that
deformation of the  membrane subjected to a small stretching tension  $\sigma>0$ scales as ${\Delta L\propto\sigma^\alpha}$,
with a non-trivial exponent $\alpha$.
 In the opposite case, $\sigma<0,\ \Delta L < 0$, the membrane undergoes a buckling transition \cite{buck}.
We obtain the critical index $\alpha$ that turns out to be different for clean and disordered cases.
Our findings imply that for sufficiently strong disorder the anomalous elasticity of
graphene is fully determined by static random deformations---ripples. The non-linearity of
elasticity of graphene found in this work is in agreement with recent experimental findings~\cite{lopez-polin15,nicholl15}.
Related theoretical results have been recently obtained for
clean membranes in the ribbon geometry~\cite{nelson15} and by numerical simulations~\cite{katsnelson16}.

We  consider a $2D$ membrane embedded in the $d$-dimensional space ($d>2$).
The starting point of our analysis is the energy functional
\be
E\!=\!\!\!\int\!\!d^2x\!\left[\!\frac{\varkappa }{2}(\Delta{\mathbf r})^2\!
+\!\frac{\mu}{4}(\partial_\alpha{\mathbf r}\partial_\beta{\mathbf r}\!-\!\delta_{\alpha\beta})^2
\!+\!\frac{\lambda}{8}(\partial_\gamma{\mathbf r}\partial_\gamma{\mathbf r}\!-\!D)^2\!\right]
\nonumber
\ee
which can be obtained from the general  gradient expansion of  elastic
energy  \cite{NelsonCrumpling}
by using a certain rescaling of coordinates (see discussion in \cite{my-crump}).
Here  $\varkappa$ is the bare bending rigidity, while $\mu$ and $\lambda$ are
in-plane coupling constants. The $d-$dimensional vector $\mathbf r= \mathbf r(\mathbf x)$ describes
a point on the membrane surface and depends on the $2D$ coordinate $\mathbf x$ that parametrizes the
membrane.
The vector $\mathbf r$ can be split into
$ \mathbf r=\xi \mathbf x+ \mathbf u +\mathbf h,  $
where  vectors
$\mathbf u=(u_1,u_2),~ \mathbf h =(h_1,...,h_{d_c}) $
represent in-plane and out-of-plane displacements, respectively, and $d_c=d-2.$
The stretching factor $\xi$ is equal to unity in the mean-field approximation but
gets reduced due to fluctuations.
In terms of $\mathbf u,\ \mathbf h$, and $\xi$, the energy becomes
\be
E\!=\!\frac{L^2(\mu\!+\!\lambda)(\xi^2\!-\!1)}{2}\!\left[\xi^2\!-\!1\!+\!\int\!\frac{d^2\mathbf x}{L^2}
\p_\alpha{\mathbf h}\p_\alpha{\mathbf h}\right]\!+\!E_0,
 \label{zero mode}
\ee
where $\tilde {\mathbf u}=\xi \mathbf u$ and  $E_0=E_0(\tilde {\mathbf u}, \mathbf h)$
describes the energy of in-plane and out-of-plane fluctuation.
We proceed now to
include the static disorder. As
shown in Ref.~\cite{my-crump},
the relevant disorder is
produced
by a random curvature. The
energy of fluctuations including such disorder reads~\cite{disorders}
\be
E_0 (\mathbf u,\mathbf h)= \int d^Dx\left\{  \frac{\varkappa}{2} (\Delta\mathbf h +\boldsymbol{\beta})^2+\mu u_{ij}^2+\frac{\lambda}{2}u_{ii}^2   \right\}.
\label{Fdis}
\ee
Here
${u_{\alpha\beta}=\left(\p_\alpha u_\beta+\p_\beta u_\alpha +\p_{\alpha}{\mathbf h}\p_\beta{\mathbf h}\right)/2}
 $
is the strain tensor and
${\boldsymbol{\beta}=\boldsymbol{\beta(\mathbf x)}}$ is a random vector with Gaussian distribution
$
{P(\boldsymbol{\beta})=Z_{\boldsymbol{\beta}}^{-1}\exp\left[ -({1}/{2b})\int \beta^2(\mathbf x) d^D\mathbf x \right]},
$
where $b$ is the disorder strength and $Z_{\boldsymbol{\beta}}$ is a normalization factor.
For $\boldsymbol \beta=0,$
$E_0 (\mathbf u,\mathbf h) $ coincides with the conventional expression for elastic energy of nearly flat membrane \cite{Nelson}.

The second term in the square brackets in  Eq.~\eqref{zero mode}
describes the coupling between fluctuations and stretching.
Such a coupling leads to shrinking of the membrane in the  $x-$plane.
As a result, the optimal value of $\xi$ deviates from the mean-field value $\xi=1$ due to the fluctuations.
The size of the membrane with fluctuations $R$  is related to the size  $L$ of the  membrane
without fluctuations as follows: $R=\xi L.$
Hence, the  ``projected'' area of the membrane reads $A=\xi^2 L^2.$

For $\sigma=0,$ the equilibrium value of $\xi$
reads \cite{my-crump,my-supp}:
  \be
  \xi^2=1-{\left \langle \p_\alpha \mathbf h  \p_\alpha \mathbf h \right \rangle}/{2}.
  \label{dxi}
\ee
Here angular brackets denote the Gibbs averaging.
Application of
tension $\sigma$ to the membrane leads to the increase of $ \xi^2$,
as compared to Eq.~\eqref{dxi}.
Below we calculate  function $\xi(\sigma)$,
both for clean and disordered cases.

{\it Clean case} ($b=0$).
For $\sigma \neq 0,$ the
propagator of
$h-$modes calculated
in the harmonic approximation is given by (see  \cite{my-supp} for technical details)
\be
\langle  h_{ \mathbf q}^{ \alpha} h_{ - \mathbf q'}^{ \beta} \rangle=  (2\pi)^2\delta(\mathbf q-\mathbf q')~\delta_{\alpha\beta}~G_\mathbf q^0,
\label{hh}
\ee
where
${G_\mathbf q^0 ={T}/({\varkappa q^4 +\sigma q^2})}.$
Tension $\sigma $  is given by a derivative of the free energy $F$ with respect to $A$ \cite{david2},
$\sigma = {\p  F}/{\p A},$
and is related to $\xi$ as \cite{my-supp}
\be
\sigma=(\mu+\lambda) \left( \xi^2-1 +{ \left \langle \p_\alpha \mathbf h  \p_\alpha \mathbf h \right \rangle}/{2} \right).
\label{sigma}
\ee
Conventional HL can be derived from Eq.~\eqref{sigma} by neglecting
the contribution of fluctuations and assuming that $\xi$ is close to unity:
${\sigma_{\rm conv}\approx k_0\left( \xi-1 \right)}.$ Here $k_0=2(\mu+\lambda) \approx 400$ N/m  is in-plane stiffness  predicted  for flat graphene \cite{jiang,shokrieh} and  measured  in  Refs.~\cite{lee,metten}.
The main purpose of the further discussion is to demonstrate that the contribution
of fluctuations is of crucial importance, so that
this law fails in the limit $\sigma \to 0$
where
stretching
turns out to be
a nonlinear function of $\sigma.$

For large momenta, ${q>q_\sigma},$  where
${q_\sigma=\sqrt{\sigma/\varkappa}},$
Green's function is approximately  given by
${G_{\mathbf q}^0={T}/{\varkappa q^4}}.$
The strong infrared singularity  ${G_{\mathbf q}^0 \propto 1/q^4}$
leads to a logarithmic divergence of
${\left\langle\p_\alpha{\mathbf h}\p_\alpha{\mathbf h}\right\rangle}$ and,
consequently, in view of Eq.~\eqref{dxi}, to the renormalization of $\xi $ \cite{my-crump}.
Hence, $\xi$ becomes scale-dependent: ${\xi\to\xi_L}$, where ${L\sim q^{-1}}$.
At finite $q,$ the renormalization is stopped because of the term $\sigma q^2$ in the denominator
of $G_\mathbf q^0.$   To determine the $q$-dependence of the renormalization of $\xi$,
one should take into account that the bending rigidity is also renormalized for sufficiently small wave vectors
$q\ll q_*$  according to the RG equation \cite{Aronovitz89,lower-cr-D2,Doussal},
\be
 {d\varkappa}/{d \Lambda}= \eta \varkappa \,\,\,\Rightarrow \,\,\,\varkappa_q=\varkappa \left( {q_*}/{q}\right)^\eta.
\label{kappa0}
 \ee
Here $\Lambda= \ln(q_*/q)$,
$\eta$ is the anomalous dimension of the bending rigidity,
$q_*$ is the inverse Ginzburg length,
\be
q_* \simeq {\sqrt{\tilde{\mu} ~T}}/{\varkappa},
\label{ginz1}
\ee
and ${\tilde{\mu}=3\mu(\mu+\lambda)/[8\pi(2\mu+\lambda)]}$,
see
Ref.~\cite{my-crump}.
Below, we assume that $q_* \gg q_\sigma.$
In this case, a competition between the two terms in the denominator of $G_\mathbf q^0$
leads to appearance of a new spatial scale $\tilde{q}_\sigma$ determined by the condition $\varkappa_q q^2 =\sigma$, yielding
$\tilde{q}_\sigma =q_\sigma  \left(  {q_\sigma}/{q_*}\right)^{\eta/(2-\eta)}.$
Next, we calculate ${\left\langle\p_\alpha{\mathbf h}\p_\alpha{\mathbf h}\right\rangle}$ with the use of
Eq.~\eqref{kappa0} and substitute it in Eq.~\eqref{sigma} (see \cite{my-supp} for details).
This yields an equation that determines the dependence of $\xi=\xi_{L\to \infty}$ on $\sigma,$
\be
\frac{\sigma}{\mu+\lambda}=\xi^2-1+ \frac{d_c T}{4\pi} \int\limits_0^{q_{\rm uv}}\frac{q dq }{\varkappa_q q^2 +\sigma},
 \label{an-hooke}
\ee
where $q_{\rm uv}$ is the ultraviolet cutoff ($q_{\rm uv} \gg q_*$).
 In the absence of stress ($\sigma=0$),  Eq.~\eqref{an-hooke} simplifies.
For ${d_c\gg 1},$ when ${\eta=2/d_c},$  one gets \cite{david1,my-crump}
\be
\xi^2|_{\sigma=0}  \equiv \xi_0^2=1-\varkappa_{\rm cr}/\varkappa=1-T/T_{\rm cr},
\label{xiT}
\ee
where ${\varkappa_{\rm cr}={d_c^2 T}/{8\pi}}$ and  ${T_{\rm cr}={4\pi \eta \varkappa}/{d_c}}$
is the temperature of crumpling transition (CT)  for a given value of bare
bending rigidity $\varkappa.$
 For $T<T_{\rm cr}$, the stretching factor is finite, $\xi_0>0$,
and the membrane is in the flat phase. For $T>T_{\rm cr}$, the membrane undergoes the
CT, so that $\xi \to 0 $
  for
$L<\infty.$
Interestingly, Eq.~\eqref{xiT} predicts a negative expansion coefficient  of the membrane, $d\xi_0/dT<0$.

For $\sigma \neq 0,$ we assume for simplicity  $ d_c=1,$ $ \mu\sim \lambda \sim k_0,$ (this is the case for graphene)  and rewrite Eq.~\eqref{an-hooke}  as follows (see derivation in  \cite{my-supp})
  \be
  \frac{ 2 \sigma_*}{k_0} \left[ \frac{\sigma}{\sigma_*} + \frac{1}{\alpha} \left(\frac{\sigma}{\sigma_*}\right)^\alpha\right]=\xi^2- \xi_0^2,
  \label{an-hooke1}
  \ee
where
\be
 \alpha=\eta/(2-\eta),\quad \sigma_* = C {  k_0 T}/{T_{\rm cr}},
\label{sigma*}
\ee
and $C \sim 1$ is a numerical coefficient.
Equation \eqref{an-hooke1} represents a general form of HL for clean membrane.
The l.h.s. of this equation contains two terms: a regular term, proportional to $\sigma$, and an irregular one that shows
a fractional scaling with $\sigma$. Analytical approximations \cite{Doussal}, as well as numerical simulations \cite{Gompper91,Bowick96} for the physical case $D=2$, $d=3$, show that $\eta \simeq 0.7$,  yielding   $\alpha\simeq 0.54 <1$.
Hence, the irregular term  dominates at small $\sigma$, and $\xi$
shows an anomalous
behavior,
while the
linear HL, $d\sigma/ d\xi=k_0$,
is realized for sufficiently large tensions, $\sigma\gg\sigma_*$.
 For sufficiently low temperatures, $T\ll T_{\rm cr}$, the  stretching  corresponding  to $\sigma_*$ is small,   $\xi_*-\xi_0\sim {T}/{T_{\rm cr}}\ll1$.
For $\sigma >\sigma_*$, the term $(\sigma/\sigma_*)^\alpha$ becomes subleading.  (In this case  $ \tilde{q}_\sigma  $ turns out to be  larger than $q_*$,   which leads to additional suppression of this term, $\alpha^{-1}(\sigma/\sigma_*)^\alpha \to \ln (\sigma/\sigma_*)$   \cite{my-supp}).
One may introduce two exponents, governing the stretching in the anomalous regime.
Far from the transition point ($T<T_{\rm cr}$),
one can expand  ${\xi^2-\xi_0^2\approx 2(\xi-\xi_0)\xi_0},$ thus finding
\be
\xi -\xi_0\propto  \sigma ^{\alpha},\,\,\,\text{far from CT point}.
\label{far-tr}
\ee
Exactly at the transition point $T=T_{\rm cr}$, $\xi_0=0$ and
\be
\xi \propto  \sigma ^{\alpha/2},\quad \text{at the CT point}  .
\label{tr}
\ee

The above results can be easily generalized to an arbitrary dimensionality of the membrane, $D>2$, by replacing
${d^2{\mathbf q}\to d^D{\mathbf q}}$ in Eq.~\eqref{an-hooke}.
This leads to the
replacement
$\alpha \to (D-2+\eta)/(2-\eta)$
of the critical index in Eqs.~\eqref{an-hooke1}, \eqref{far-tr}  and \eqref{tr} .
The latter  equation for $\alpha$  was obtained previously in Refs.~\cite{buck, david2} for $\eta=0$,
which corresponds to the case $d_c=\infty$  \cite{comm}, and predicted in \cite{lower-cr-D2}
from scaling considerations.
As seen from Eq.~\eqref{an-hooke1}, the tension leads to an increase of $T_{\rm cr}$ and,
respectively, to a decrease of $\varkappa_{\rm cr}$.
Indeed, setting $\xi=0$  in Eq.~\eqref{an-hooke1} and assuming that $\sigma \ll \sigma_*$,
we find the tension-induced change of the critical temperature,
$
{{\delta T_{\rm cr}}/{T_{\rm cr}}=-{\delta \varkappa_{\rm cr}}/{\varkappa_{\rm cr}}\sim\left({\sigma}/{k_0}\right)^{\alpha}}.
$

{\it Disordered case.}
The derivation of  perturbative RG equations for disordered graphene is performed
by using replica trick within  RPA scheme, in analogy with the case  $\sigma=0$ studied in Ref.~\cite{my-crump}.
Technical details of calculations are presented in \cite{my-supp}.
First, we  find
${\overline{\left\langle\p_\alpha{\mathbf h}\p_\alpha{\mathbf h}\right\rangle}}$ in the harmonic approximation:
 \be
 \overline{\left \langle \p_\alpha \mathbf h  \p_\alpha \mathbf h \right \rangle}= \int \frac{d^2\mathbf q}{(2\pi)^2} \frac{d_c~ T}{\varkappa q^2+\sigma}\left[  1+f\frac{\varkappa q^2}{\varkappa q^2+ \sigma}\right].
  \label{corr}
  \ee
Here
the overbar
denotes the disorder averaging and
${f={b\varkappa}/{T}}$
is a dimensionless parameter  characterising the ratio of disorder to thermal fluctuations.
For fixed $\varkappa$ and $f,$ the integral in Eq.~(\ref{corr}) logarithmically diverges
for  $\varkappa q^2 \gg \sigma$ and saturates for $\varkappa q^2 \ll \sigma.$
In view of Eq.~\eqref{dxi}, we conclude that  $\xi$ is renormalized:
\be
\frac{d {\xi}^2}{ d\Lambda} \approx - \frac{{d}_c}{4\pi}\frac{T}{\varkappa}\left(1 + f \right),
\quad \text{for} \,\,  q \gg \tilde{q}_\sigma  ,
\label{RGxixi2D}
\ee
 and $d {\xi}^2/ d\Lambda =0$ for $q \ll  \tilde{q}_\sigma.$
  The Ginzburg scale $q_*$
is also affected by disorder \cite{my-crump}:
${q_*\sim\sqrt{{\tilde{\mu}T(1+2f)}}/{\varkappa}}.$
For strong disorder or low temperatures, $f\gg 1,$ we find that
\be
q_*\sim  \sqrt {{\tilde{\mu} b}/{\varkappa}}
\label{q*disdis}
\ee
is independent of temperature, while for weak disorder ($f\ll 1$), we recover Eq.~(\ref{ginz1}), $q_*\propto T^{1/2}$.
Below we show that $\tilde{q}_\sigma$ is also modified by sufficiently strong disorder.

In the harmonic  approximation, $\varkappa$ and $f$ are constants.
However, they become scale-dependent due to the coupling between
in-plane and out-of-plane fluctuations: $\varkappa\to\varkappa_q$ and $f \to f_q.$
For $q\gg \tilde{q}_\sigma,$ corresponding RG equations were derived in  Ref.~\cite{my-crump} [see also Eq.~(S38) of \cite{my-supp}].
For strong disorder, ${f\gg 1},$ the RG equations look:
${{d\varkappa}/{d\Lambda}={\eta \varkappa}/{4}}$ and ${{df}/{d\Lambda}=-{3\eta}/{4}}$.
The first equation yields ${\varkappa_q=\varkappa (q_*/q)^{\eta/4}}$.
Equating $\varkappa_q q^2$ to $\sigma$, we find:
\be
\tilde{q}_\sigma=q_\sigma\left({q_\sigma}/{q_*}\right)^{\eta/(8-\eta)},
\label{q-sigma-dis}
\ee
where $q_*$ is given by Eq.~\eqref{q*disdis}.
Since $\varkappa $ changes faster than $f,$ one can set ${f={\rm const}}$ in Eq.~\eqref{RGxixi2D}.
Using Eq.~\eqref{sigma}, we find that the equation that determines the
dependence of $\xi$ on $\sigma$ for a strongly disordered membrane
is given by  Eq.~\eqref{an-hooke1} with   $\xi_0^2=1-B,$
\be\alpha=\eta/(8-\eta)\simeq 0.1 ,\quad\text{and} \quad  \sigma_* = C' k_0 B,
\label{sigma*dis}
\ee
where ${B=bd_c^2/2\pi}$ and $C' \sim 1$ is a numerical coefficient.
Note
that the temperature drops out from the Hooke's law for disordered membrane.
For ${\sigma=0},$
the CT (${\xi=0}$)
corresponds to ${B=B_{\rm cr}=1}$, in agreement with previous study (see Fig.~5 of Ref.~\cite{my-crump}). For $B\ll B_{\rm cr}$ the stretching corresponding to $\sigma_*$  reads $\xi_*-\xi_0 \sim B/\alpha.$
The tension enhances the critical value of disorder:
  ${\delta B_{\rm cr}\equiv B_{\rm cr}-1\sim \alpha^{-1}\left({\sigma}/{{k_0}}\right)^{\alpha}}.$
The anomalous stress-strain relations have the form \eqref{far-tr} and \eqref{tr} for $B<B_{\rm cr}$ and $B=B_{\rm cr}, $ respectively, with  appropriate  replacement of $\alpha$: the ``clean'' value (\ref{sigma*}) is replaced by the considerably smaller ``dirty'' value (\ref{sigma*dis}).
Hence, stretching of a strongly disordered membrane is a nonlinear function of a weak tension,
just as in the clean case. However, the corresponding power-law exponents differ from that of
a clean system. As in the clean case,
the conventional HL is restored for $\sigma \gg \sigma_*.$

\begin{figure}[t]
\centerline{\includegraphics[width=0.4\textwidth]{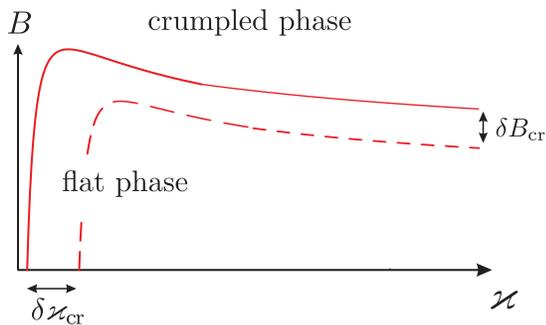} }
\caption{
  Phase diagram of graphene in the plane of parameters $\varkappa$ (bending rigitidy) and $B$ (disorder) at non-zero tension $\sigma$.
   The CT separating crumpled and flat phases is shown by full red line;
  dashed line represents the  CT for $\sigma =0$
 \cite{my-crump}.
    }
\label{F1}
\end{figure}

The RG flow for $\xi$ stops at ${q\sim\tilde{q}_\sigma}$. Thus, if $\xi>0$ at this scale, the system is in the flat phase. Conversely, if $\xi$ becomes zero before $\tilde{q}_\sigma$ is reached, the membrane crumples.
The phase diagram in the parameter plane  $(\varkappa,B)$, as obtained by numerical solution of RG equations,
is shown in Fig.~\ref{F1}. The tension shifts the line separating the flat and crumpled phases; this shift is characterized
by $\delta \varkappa_{\rm cr}$ and $\delta B_{\rm cr}$.
Interestingly, the RG flows for $\varkappa $ and $f$ do not stop at the point ${q\sim \tilde{q}_\sigma}$  \cite{my-supp}. However, for smaller $q$, such that ${\varkappa_qq^2\ll\sigma}$,  the scaling of $\varkappa$ is irrelevant for the CT and the membrane remains flat.

Both in the clean and disordered case,
it is convenient to
introduce the effective stiffness
\be
 k_{\rm eff}=\p\sigma/\p\xi \simeq k_0\frac{(\sigma/\sigma_*)^{1-\alpha}}{1+(\sigma/\sigma_*)^{1-\alpha} }.
 \label{k-eff}
\ee
 It is strongly reduced for a weak strain ($\sigma\ll\sigma_*$),
vanishing at the point of the buckling transition ($\sigma=0$).

Let us now discuss characteristic values of parameters for the case of graphene.
In Ref.~\cite{my-crump} we estimated the amplitude of the static disorder
as $b=0.03$  based on experimental measurements of parameters of ripples~\cite{Kirilenko}.
Taking the bare value of the bending rigidity for graphene, ${\varkappa\simeq 1}$~eV, we find
${f\simeq 1}$ at room temperature. This implies that at ${T\simeq 300}\,$K
the system is in the crossover regime between the clean and disordered limits. In this regime, the exponent
$\alpha$ takes a non-universal value between the  clean ($\alpha \simeq 0.5$) and disordered  ($\alpha \simeq 0.1$)  values. For low
low temperatures, ${T\ll 300}\,$K, we predict then the disordered value $\alpha \simeq 0.1$, while for elevated temperatures the clean value $\alpha \simeq 0.5$ should be reached. (In fact, $\alpha$ flows as a function of $\sigma$, tending to the clean value $\simeq 0.5$
for smallest strains. This flow is, however, logarithmically slow and may be difficult to observe experimentally.) Clearly, the crossover temperature may vary depending on sample preparation (degree of disorder).
For clean samples, we estimate the crossover tension and stretching at ${T\simeq 300}\,$K from Eq.~ \eqref{sigma*}, yielding  $\sigma_* \simeq  1\,$N/m  and $\xi_*-\xi_0\simeq 0.003$ (for $\eta=0.7$ and $C=1$). For disorder-dominated samples with the above disorder strength $b=0.03$, we get $B\simeq0.005$, which yields, according to \eqref{sigma*dis}, an estimate $\sigma_* \simeq  2\,$N/m and $\xi_*-\xi_0\simeq 0.05$ (for $C'=1$).

\begin{figure}[t]
\centerline{\includegraphics[width=0.5\textwidth]{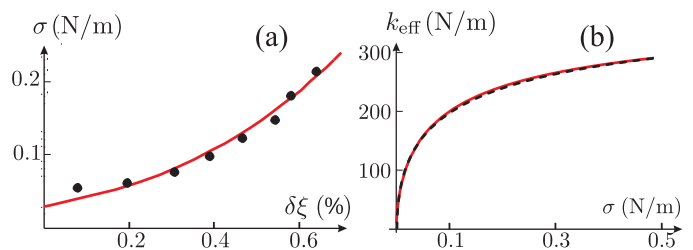} }
\caption{ (a) Stress-strain dependence. Dots --  experiment
\cite{nicholl15},  line -- theory for strongly disordered case $\alpha=0.1 $ with degree of disorder $B=0.004$. (b) Effective stiffness $k_{\rm eff}$ {\it vs.} stress $\sigma$ in clean graphene at $T=300$K.  Dashed line -- numerical simulations \cite{katsnelson16},  solid line -- Eq.~\eqref{k-eff} with $\alpha=0.62$ (i.e., $\eta=0.765$) and
 $\sigma_* \simeq  0.1\,$N/m.    }
\label{F2}
\end{figure}

Our results compare well with a recent detailed experimental study of graphene elasticity \cite{nicholl15}.  It was found there that the room-temperature  in-plane stiffness of graphene is reduced compared to its value $k_0$  for ``ideal'' graphene (no disorder, ${T=0}$) by a large factor (up to $\sim 40$) at low stretching.  When temperature was lowered down to 10~K, the stiffness showed a sizeable increase, still remaining much smaller than 400 N/m.  These data are in agreement with our conclusion that ripples and FP strongly weaken the in-plane stiffness,
  yielding comparable contributions at room temperature.
  In Fig.~\ref{F2}a  we  compare our theory with the  strain-stress dependence presented in  Fig. 2c of Ref.~\cite{nicholl15}.        We use Eq.~\eqref{an-hooke1} describing both clean and disorder case with the appropriate choice of   $\alpha$ and $\sigma_*$, considering  $\alpha$ and $\sigma_*$ as fitting parameters. The solid lines in In Fig.~\ref{F2}a show dependence of $\sigma $ on $\delta \xi=\xi(\sigma) - \xi(\sigma_0)$,  where $\sigma_0$ is built-in stress extracted from experimental data \cite{nicholl15}.  The best fit is achieved for
   $\alpha \approx 0.1$ and   $\sigma_* \approx 1.68 $ N/m.   The obtained value of $\alpha$ implies  that  the sample is in the disorder-dominated regime, with the degree of disorder $B \simeq 0.004$  (we set  the numerical coefficient $C'=1$ here).  This estimate is in  agreement with the value  $B=0.005$    \cite{my-crump} obtained  from the transmission-electron-microscopy data of Refs.~\cite{Meyer,Kirilenko}. For a more detailed comparison with experiment, measurements  of strain-stress curves in a wide range of temperatures would be of great interest.

Our results also compare very well with numerical simulations of Ref.~\cite{katsnelson16}
which were performed for clean graphene.
In particular, the scaling of $k_{\rm eff}$  for $f\ll 1$
is in an excellent  agreement with the large-sample data (number of atoms 37888).
 For comparison, we used the empirical formula (11) of Ref.~\cite{katsnelson16} which perfectly fits numerical data obtained there (see \cite{my-supp} for details).
As seen from Fig.~\ref{F2}b, the numerical data are very well described
by our Eq.~(\ref{k-eff}) with $\alpha\simeq 0.62$ ($\eta \simeq 0.76$), as expected in the clean limit.  The comparison to the results of numerical simulations allows us to determine the numerical coefficient $C$ in Eq.~\eqref{sigma*}, which turns out to be $C\approx0.093.$  The corresponding crossover stress and strain values read
$\sigma_* \simeq  0.1\,$N/m   and $\xi_*-\xi_0\simeq 0.0006$, respectively.

To conclude, the theory of anomalous Hooke's law has been developed, for both clean and disordered graphene.
In both cases, scaling of the deformation with the external force obeys a fractal power law
in the limit of weak forces.
This behavior
is dominated by thermal fluctuations for clean graphene, while
for strongly disordered graphene
it is governed by static ripples.
Remarkably, the same coupling between longitudinal and transverse modes that enhances the bending rigiditiy,
thus rescuing the flat phase of the membrane,  leads simultaneously to a dramatic softening of the in-plane elasticity.

We thank  M.I. Katsnelson for useful comments, and K. Bolotin and  R. Nicholl  for  providing us with experimental data of Ref. ~\cite{nicholl15} and stimulating discussions. The work was supported by the Russian Science Foundation (grant No. 14-42-00044).

\newpage
\mbox{}
\newpage
\begin{center}
\textbf{SUPPLEMENTAL MATERIAL}
\end{center}

\setcounter{equation}{0}
\setcounter{figure}{0}
\renewcommand{\theequation}{S.\arabic{equation}}
\renewcommand{\thefigure}{S.\arabic{figure}}
\renewcommand{\cite}[1]{{[}\onlinecite{#1}{]}}

\renewcommand{\thepage}{S\arabic{page}}
\renewcommand{\thesection}{S\arabic{section}}
\renewcommand{\theequation}{S.\arabic{equation}}
\renewcommand{\thefigure}{S\arabic{figure}}
\renewcommand{\bibnumfmt}[1]{[S#1]}

\setcounter{page}{1}
\setcounter{section}{0}
\setcounter{equation}{0}
\setcounter{figure}{0}

\subsection{Derivation of  the  free energy of the clean  membrane with nonzero tension}

We start with Eq.~(1) of the main text for the elastic energy (in the absence of disorder, $\boldsymbol{\beta}=0$) and rewrite it as  follows
\be
E=\frac{ L^2 (\mu+\lambda )}{2} \left[ \left(\xi^2-1+\frac{ K  }{2} \right)^2   - \frac{K^2}{4} \right] + E_0(\tilde{ \mathbf u}, \mathbf h),
       \label{zeromode}
   \ee
where
\be
K= \int \frac{d^2x}{L^2}  \p_\alpha \mathbf h  \p_\alpha \mathbf h
\label{K}
\ee
and $E_0(\tilde{ \mathbf u}, \mathbf h)$ is given by Eq.~(2) of the main text (with $\boldsymbol{\beta}=0$).
Next, we  calculate the free energy $F$ from the partition function $Z$ that
is written as a functional integral over fluctuations  $\tilde{ \mathbf u}, \mathbf h:$
\BEA  F&=&-T \ln Z, \\
Z&=&\int e^{-E/T}\{d\tilde{ \mathbf u} d\mathbf h\} \label{Z}.
 \EEA

As a first step, one can integrate out the in-plane modes $\tilde{ \mathbf u}$, thus obtaining the energy functional
that depends on $\mathbf h$ fields only \cite{Doussal,my-crump}.
The interaction between these fields is described by the quartic term
$R_\mathbf q (\mathbf k, \mathbf k') (\mathbf h_{\mathbf k+ \mathbf q } \mathbf h_{-\mathbf k})(\mathbf h_{-\mathbf k'- \mathbf q } \mathbf h_{\mathbf k'}) $ (the explicit form of $R_\mathbf q (\mathbf k, \mathbf k')$ can be found in Refs.~\cite{Doussal,my-crump}).
This interaction can be taken into account within the RPA approach.
It is screened by polarization bubbles  and  leads to the renormalization of the bending rigidity:
$\varkappa \to \varkappa_q.$  The term with $q=0$ (zero-mode) needs special attention.

There are two zero-mode contributions: the term
\BEA
E_{ZM}^{(2)}&=&-\frac{ L^2 (\mu + \lambda )K^2}{8}
\nonumber
\\
&=&-\frac{\mu+ \lambda}{8 L^2}\left[\int (dQ) Q^2(\mathbf h_{-\mathbf Q}\mathbf h_{\mathbf Q})\right]^2
\label{EZM2}
\EEA
[see Eq.~\eqref{zeromode}], and the term with $q=0$ coming from the $h^4$-terms in $E_0(\tilde{ \mathbf u}, \mathbf h)$:
\BEA
E_{ZM}^{(4)}&=&\int \frac{(dQ)(dQ')}{L^2}\left[ \frac{\mu}{4} (\mathbf Q \mathbf Q')^2 + \frac{\lambda}{8} Q^2Q'^2 \right]
\nonumber \\
&\times&
 (\mathbf h_{-\mathbf Q}\mathbf h_{Q}) (\mathbf h_{-\mathbf Q'}\mathbf h_{\mathbf Q'})
 \label{EZM4}
\EEA
[here $(dQ)=d^2\mathbf Q/(2\pi)^2$]. Note that the $h^4$-contribution arising after integrating terms of the type $\tilde u h^2$
over $\{d \tilde u\}$ does  not contain the $q=0$ term because the zero mode of the in-plane fluctuations,  $\xi \mathbf x,$
was separated from the very  beginning.
Combining the contributions (\ref{EZM2}) and (\ref{EZM4}), we find
\BEA
E_{ZM} &=&\frac{\mu}{4 L^2} \int (dQ)(dQ')  \left[ (\mathbf Q \mathbf Q')^2 -\frac{Q^2Q'^2}{2} \right]
\nonumber
 \\
&\times&
(\mathbf h_{-\mathbf Q}\mathbf h_{Q}) (\mathbf h_{-\mathbf Q'}\mathbf h_{\mathbf Q'}).
\label{EZM}
\EEA
One can calculate the ``Hartree'' and ``Fock'' contributions to the self-energy coming from the functional \eqref{EZM}.
The  Hartree  contribution vanishes after averaging over the angle between $\mathbf Q$ and $\mathbf Q'.$
The Fock contribution comes from $\mathbf Q=\mathbf Q'$ and hence survives the angular averaging.
The lowest-order Fock correction to the self-energy is momentum-independent, but is inversely proportional to the system size:
$\Sigma \propto \varkappa  q_*^2/L^2,$ thus vanishing in the thermodynamic limit.
Taking screening into account leads to a further suppression of the Fock self-energy.
Indeed, the inverse polarization operator increases with decreasing $q:$ $\Pi_\mathbf q^{-1}  \propto q^2$,
see Eq.~(38) of Ref.~\cite{my-crump}.  Since the interaction line in the zero-mode terms
carries zero momentum,  we take polarization operator at $q \sim 1/L$ and finally obtain:
$\Sigma \propto \varkappa  /L^4.$
Thus, the zero-mode interaction can be safely neglected.
It is worth noting that the key point of this derivation is the cancellation of the Hartree contribution
after the angular averaging.
Indeed, one can check that each of the two terms  of the opposite sign in Eq.~\eqref{EZM} yields a correction to self-energy in the Hartree channel
which does not depend on the system volume.

We are, therefore, left with the  following effective functional
 \be
 E=\frac{L^2(\mu +\lambda)}{2}\left(\xi^2-1+\frac{K}{2}\right)^2+ \int (dq) \frac{\varkappa_q}{2} q^4 h_\mathbf q h_{-\mathbf q}.
 \label{Eh}
 \ee
Since the in-plane modes are integrated out, the partition function, Eq.~\eqref{Z}, contains now the integral over $\{d\mathbf h\}$ only.
To do this integration, we  first introduce an integral  over an auxiliary field $\chi, $
  \BEA
 && \exp \left[  -\frac{L^2(\mu+\lambda)}{2T}\left(\xi^2 -1+\frac{K}{2}\right)^2 \right] = \frac{L}{\sqrt{2\pi(\mu+\lambda) T}}
  \nonumber
 \\
 \nonumber
 && \times \int d\chi \exp\left\{-\left[\frac{\chi^2 }{2(\mu+\lambda)} - i \chi \left(\xi^2-1+\frac{K}{2}\right)\right]\frac{L^2}{T}\right\}.
 \EEA
Next, we calculate the Gaussian integral over $\{d\mathbf h\}$ and get (omitting irrelevant constants)
\be
F=-T \ln\left[ \int d\chi \exp\left( -\frac{L^2 S}{T}\right)\right],
\label{F}
\ee
where
\BEA
S=S(\chi,\xi)
&=&\frac{\chi^2 }{2(\mu+\lambda)} - i \chi (\xi^2-1)\nonumber \\
&+& \frac{d_c T}{2}\int (dq) \ln(\varkappa_q q^2 -i\chi).
\label{Schixi}
\EEA
The stationary phase  condition, $\p S/\p \chi=0,$ for the integral in Eq.~\eqref{F} yields
\be
\chi=\chi_0=i \sigma,
\ee
where $\sigma$ is related to $\xi$  by Eq.~(8) of the main text.
Calculating the integral over $d\chi$, we find
\be
F= -\frac{\sigma^2 L^2}{2(\mu+\lambda)} +\sigma L^2(\xi^2-1) +\frac{d_cT}{2}\sum_\mathbf q \ln(\sigma+\varkappa_q q^2).
\label{Fxi}
\ee
Here we omitted  terms independent on $\xi$ and $\sigma$ as well as  terms, which are small with respect to the system size.
%
Differentiating  Eq.~\eqref{Fxi} with respect to the projected area $A=\xi^2L^2$ yields
$\sigma=\p F/\p A$ as it should be.
For $\sigma \neq 0,$ the Green function of the out-of-plane modes reads
\be
\langle h_\mathbf q^\alpha   h_{-\mathbf q'}^\beta \rangle   = (2\pi)^2 \delta_{\alpha\beta}~\delta(\mathbf q-\mathbf q') \frac{T}{\varkappa_q q^4+\sigma q^2} .
\label{hhsigma}
\ee
Here the angular brackets $\langle \cdots \rangle$ denote the  Gibbs averaging with the weight  $Z^{-1}\exp(-E/T) \{d\mathbf h\}$ and $E$ is given by Eq.~\eqref{Eh}.

\subsection{Derivation of Eq.~(10) of the main text}
Rewriting integral in the r.h.s.  of Eq.~(8) as follows
  \be
     \int_0^{q_{\rm uv}}
 \frac{ q dq}{\varkappa_q q^2 \!+\!\sigma \! }=   \int_0^{q_{\rm uv}}
 \frac{ dq}{\varkappa_q q \! } -    \int_0^{q_{\rm uv}}
 \frac{ \sigma dq}{(\varkappa_q q^2 \!+\!\sigma \!)\varkappa_q q  }
  \ee
we get
\be
\frac{\sigma}{ \mu+\lambda}\!=\!\xi^2-\xi_0^2\!
 -\frac { T}{4\pi}   \int_0^\infty dq
 \frac{ \sigma }{(\varkappa_q q^2 \!+\!\sigma \! )\varkappa_q q }   ,
\label{xi1}
\ee
(having in mind  apply our theory to graphene, we put $d_c=1$ here).
The main contribution to this  integral  comes from small $q$  that allowed us to extend upper limit of  integration to infinity.
Next, we  interpolate the bending rigidity as
\be
\varkappa_q = \varkappa\left( \frac{q+q_*}{q}\right)^\eta.
\ee
Assuming that $T\ll T_{\rm cr}$ and, consequently,   $\xi- \xi_0\ll \xi_0 \simeq 1,$
we arrive after simple algebra  at the following equation
\be
 \xi-\xi_0=  \frac{\sigma}{k_0}  + \frac{T}{8\pi \varkappa}F\left(\frac{\sigma}{\sigma_1}\right),
 \ee
 where
 \be
\sigma_1= \varkappa q_*^2 \sim  \tilde \mu T/\varkappa ,
\label{sigma1}
\ee
 and
 \BEA
 F(x)&=&   x   \int_0^\infty
 \frac{ dz}{z^{1-\eta} (1+z)^\eta \left[ z^{2-\eta} (1+z)^\eta+ x \right]  }
 \nonumber
  \\
  &\propto& \left\{    \begin{array}{c}
                           x^{\eta/(2-\eta)},  \quad \text{for}\quad  x\to 0,  \\
                          \ln x,  \quad \text{for} \quad x\to \infty.
                       \end{array}
    \right.
 \label{F}
 \EEA
 Since the numerical coefficient in Eq.~\eqref{sigma1} is unknown, we can rewrite  the low-stress asymptotics of $F(x)$ in term of $\sigma_*$  which is given by Eq.~(11) of the main text and contains an unknown temperature-independent numerical coefficient $C \sim 1$. Hence,  for $\sigma \ll \sigma_*$ we arrive at Eq.~(10) of the main text, where in addition to the main anomalous contribution $\propto \sigma^\alpha$ we keep the subleading  linear-in-$\sigma$ term. For $\sigma \gg \sigma_0$, the linear term dominates.  According to Eq.~\eqref{F},  the  anomalous term is proportional to $\ln (\sigma/\sigma_*)$ in this region. However, since the anomalous term is subleading, one can use Eq.~(10)  of the main text in this region, too, if one is only interested in describing the leading behavior.


\subsection{Renormalization group for  disordered membrane with non-zero tension}

For disordered case, one can perform calculations by using the replica trick.
Replicating the fields $h_\mathbf q\to h_\mathbf q^{(n)}$ in the energy functional
and omitting  irrelevant constant,  we obtain (see also Ref.~\cite{my-crump}):
 \be
   E^{\rm{rep}}= E^{\rm{rep}}_0+ E^{\rm{rep}}_1+E^{\rm{rep}}_2,
   \ee
   where
   \BEA
E^{\rm{rep}}_0&=&
 \frac{1}{2}\sum\limits_{n=1}^{n=N}  \int (dq) \varkappa q^4
   |\mathbf h_\mathbf q^{(n)} +\boldsymbol{\beta}_\mathbf q  |^2,
\\
E^{\rm{rep}}_1&=&\frac{L^2(\mu+\lambda)}{2}\sum\limits_{n=1}^{n=N}\left( \xi^2-1+\frac{K_n}{2}\right)^2,
\label{E1}
\\
 E^{\rm{rep}}_2 &=& \sum\limits_{n=1}^{n=N} \frac{1}{4d_c} \label{F-repl}\\
 &&\!\!\!\!\!\!\!\!\!\!\!\!\!\!\!
 \times \int (dk dk' dq) R_\mathbf q(\mathbf k, \mathbf k') \left( \mathbf h_{\mathbf k+\mathbf q}^{(n)} \mathbf  h_{-\mathbf k}^{(n)} \right)
 \left( \mathbf h_{-\mathbf k'-\mathbf q}^{(n)} \mathbf  h_{\mathbf k'}^{(n)} \right),
\nonumber
\\
K_n&=& \int \frac{d^2x}{L^2}  \p_\alpha \mathbf h^{(n)}  \p_\alpha \mathbf h^{(n)},
\label{Kn}
\EEA
and index $n=1, \ldots, N $ enumerates replicas (the rule of summation over repeated indices does not apply here).
Next, we introduce  auxiliary fields $\chi_n $ to decouple the zero-mode interaction in Eq.~\eqref{E1}
and average $\exp(-E_{\rm{rep}}/T)$ with $P(\boldsymbol{\beta} ).$
In the absence of  interaction $E^{\rm{rep}}_2,$  the stationary phase conditions  yield $\chi_1=\ldots=\chi_N=i\sigma, $
where tension $\sigma$ and stretching factor $\xi$ are connected by the following  equation:
 \be
\frac{\sigma}{\mu+\lambda}=\xi^2-1+ \frac{d_c T}{4\pi} {\rm Tr}\left( \int\limits_0^{q_*}\frac{q dq }{\hat \varkappa q^2 +\sigma}\right).
\label{an-dis}
\ee
Here  we have introduced a replica-space matrix $\hat {\varkappa}$:
\be
{\hat \varkappa}=\varkappa   - \frac{b\varkappa^2 }{T+b\varkappa N} \hat J,
\label{kappa-matrix}
\ee
where  $\hat J $ is  the  matrix  with  all elements equal to unity: $J^{nm}=1$.
It is convenient to incorporate $\sigma$ in the definition of the bending rigidity matrix by introducing
\be
{\hat \varkappa_\mathbf q}=\varkappa +\frac{\sigma}{q^2}  - \frac{b\varkappa^2 }{T+b\varkappa N} \hat J.
\label{kappa-matrix-J}
\ee
The bare propagator is then a matrix in the replica space:
\BEA
\hat G_\mathbf q^0  &=& \frac{T \hat \varkappa_\mathbf q^{-1}}{ q^4} =\frac{T  }{ \bar  \varkappa_\mathbf q  q^4}\left( 1+  \bar f_\mathbf q  \hat J \right) ,
\label{G0}
\\
\bar \varkappa_\mathbf q&=&\varkappa \frac{q^2+q_\sigma^2}{q^2},\quad
\bar f_\mathbf q=f \frac{q^2}{q^2+q_\sigma^2}.
\label{f-kappa}
\EEA
Here we have introduced the dimensionless parameter \cite{my-crump}
\be
f=\frac{b \varkappa}{ T }
\ee
given by the ratio of the bare disorder, $b,$ and the bare magnitude of
dynamical (thermal) fluctuations, $T/\kappa.$
The correlation function ${\overline{\left\langle\p_\alpha{\mathbf h}\p_\alpha{\mathbf h}\right\rangle}}$ in the harmonic approximation  is given by $N^{-1} {\rm Tr}  \hat G_\mathbf q^0.$  (it worth noting that  ${\overline{\left\langle\p_\alpha{\mathbf h}\p_\alpha{\mathbf h}\right\rangle}}$ can be calculated without replication, by the   Gibbs averaging of  $\p_\alpha{\mathbf h}\p_\alpha{\mathbf h}$  with the fluctuation energy $E_0,$ where one should omit anharmonic terms).

Above we have neglected the anharmonic coupling, $E^{\rm{rep}}_2 ,$ so that  $\varkappa$ and $b$ were $q$-independent.
In fact, the anharmonicity leads to a weak scale dependence of these variables:
$\varkappa \to \varkappa_\mathbf q, ~b \to b_\mathbf q,~f \to f_\mathbf q $.  To find this dependence, we calculate
the anharmonicity-induced  self-energy $\hat \Sigma_\mathbf q$  and find the dressed Green function
\be
\hat G_\mathbf q= \frac{T}{\hat \varkappa_\mathbf q q^4 +\hat \Sigma_\mathbf q }.
\ee

Next, we derive the RG equation for $\hat \Sigma_\mathbf q$  at the one-loop order (i.e., within RPA). The analysis is controlled by a parameter $1/d_c$ which is assumed to be small. We first calculate the polarization operator which
  becomes a replica-space matrix \cite{my-crump}:
\begin{eqnarray}
\label{Pinm}
  \Pi_\mathbf q^{nm} &=& \frac{1}{3} \int (dk) k_\perp^4 G_\mathbf k^{0,nm} G^{0,nm}_{\mathbf q-\mathbf k} \\
   &=& \frac{T^2}{3} \int (dk) k_\perp^4
   \frac{(\hat\varkappa_\mathbf k^{-1})^{nm}}{k^4}\frac{(\hat\varkappa_{\mathbf q-\mathbf k}^{-1})^{nm}}{|\mathbf q-\mathbf k|^4} .
 \nonumber
\end{eqnarray}
Here
$\mathbf k_\perp= \hat P \mathbf k=\mathbf k-\mathbf q (\mathbf k\mathbf q)/q^2,$
and $\hat P$ is the  projection  operator related to the transferred momentum $\mathbf q$,
\be
P_{\alpha\beta} =\delta_{\alpha\beta}- q_\alpha q_\beta/q^2.
\ee
From Eqs.~\eqref{G0} and \eqref{Pinm} we find
\be
\label{Pinm-new}
  \hat \Pi_\mathbf q
   = \frac{T^2}{3} \int (dk) k_\perp^4
    \frac{1+\bar f_\mathbf k+\bar f_{\mathbf q-\mathbf k}
   +\hat J \bar f_\mathbf k \bar f_{\mathbf q-\mathbf k} }{\bar \varkappa_\mathbf k k^4 \bar \varkappa_{\mathbf q-\mathbf k} |\mathbf q-\mathbf k|^4}.
 \ee
 Using Eq.~\eqref{f-kappa} we find that behavior of $\hat \Pi_\mathbf q$ is essentially different for $q\gg q_\sigma$ and $q \ll q_\sigma:$
\be
 \hat \Pi_\mathbf q =\frac{T^2}{16\pi\varkappa ^2} \left\{  \begin{array}{c}
                                 \displaystyle \frac{1}{q^2} \left(1+2f +f^2 \hat J\right), \,\,\,\text{for}\,\, q \gg q_\sigma, \\[0.5cm]
                                   \displaystyle \frac{1}{2 q_\sigma^{2}}\left(1 +  f +\frac{f^2}{12} \hat J\right) , \,\,\,\text{for}\,\, q \ll q_\sigma,
                               \end{array}
\right.
  \label{Pi0nm}
 \ee
The upper ($q\gg q_\sigma$) asymptotics in Eq.~(\ref{Pi0nm}) has been obtained previously in Ref.~\cite{my-crump}.
We assume that $q \ll q_*$ (relation between $q$ and $q_\sigma$ can be arbitrary), where the Ginzburg scale $q_*$ is
modified by disorder \cite{my-crump}:
 \begin{equation}
q^* \sim\left[\frac{\tilde{\mu} T (1+2f)}{\varkappa^2}\right]^{1/2}.
\label{q*dis}
\end{equation}
It is worth noting that for strong disorder or low temperatures ($f\gg 1$) $q_*\sim (\tilde{\mu} b/\kappa)^{1/2}$
is independent of temperature, while for weak disorder ($f\ll 1$) , $q_*\propto T^{1/2}$.
For such $q,$  matrix elements  of the self-energy in the replica space read
\be
\Sigma_\mathbf k^{nm}=\frac{2T}{3 d_c}\int (dq) k_\perp^4  { \left(\hat \Pi_\mathbf q^{-1}\right)^{nm}}G_{\mathbf k-\mathbf q}^{0,nm}.
\label{Sigmanm}
\ee
Substituting here the bare Green function, Eq.~\eqref{G0},  we find  by simple power-counting that
the corresponding integral diverges as $ k^4\ln k$ both for $k>q_\sigma$ and $k<q_\sigma.$
This implies that RG equations can be written in terms of the renormalization of $\hat \varkappa.$
Separating in thus obtained equation terms proportional to unity and to $\hat J$, we arrive to
\be
\label{dkappa2-df2}
    \frac{1}{\varkappa} \frac{d \varkappa }{d\Lambda}= \eta    \frac{1+3f+f^2}{(1+2f)^2},  \quad  \frac{1}{f}\frac{d  f }{d\Lambda}= -\eta      \frac{1+3f}{(1+2f)^2}.
     \ee

\subsection{Dynamical and static fluctuations in the flat phase. Flat phase 1 and flat phase 2}

Here, we analyze the RG flows for $\varkappa $ and $f$  at  ${q< \tilde{q}_\sigma.}$
The corresponding RG equations are analogous to the case $\sigma=0$, yielding
for ${q\ll\tilde{q}_\sigma}$
\be
\label{dkappa-df-4}
    \frac{1}{\varkappa} \frac{d \varkappa }{d\Lambda}=  -\frac{1}{f} \frac{d f }{d\Lambda}= \eta    \frac{12+12f-f^2}{6(1+f)^2}.
     \ee
Note that ${\varkappa f={\rm const}}$. Equations \eqref{dkappa-df-4} have an unstable fixed
point ${f_{\rm cr}=6+4\sqrt{3}\approx 12.9}.$
Thus, the flat phase can be separated into two parts: a phase where
$\varkappa$
increases with the system size (${f<f_{\rm cr}}$) and a phase where the membrane becomes softer
at larger scales (${f>f_{\rm cr}}$).
The border between these phases can be found by using the fact that $f_{\rm cr}$
is numerically large.
For ${q \gg \tilde{q}_\sigma}$,
using the
large-$f$ asymptotic of the RG equation for $f$,
we find that the border is given by the line
${f=f_0={\rm const}}$, where
${f_0=f_{\rm cr}+({3}/{4})\ln\left({q_*}/{\tilde{q}_\sigma}\right)}$, with
$q_*$ and $\tilde{q}_\sigma$ given by
Eqs.~(16)
 and  (18) of the main text,
 respectively, see Fig.~S1.

\begin{figure}[t]
\centerline{\includegraphics[width=0.4\textwidth]{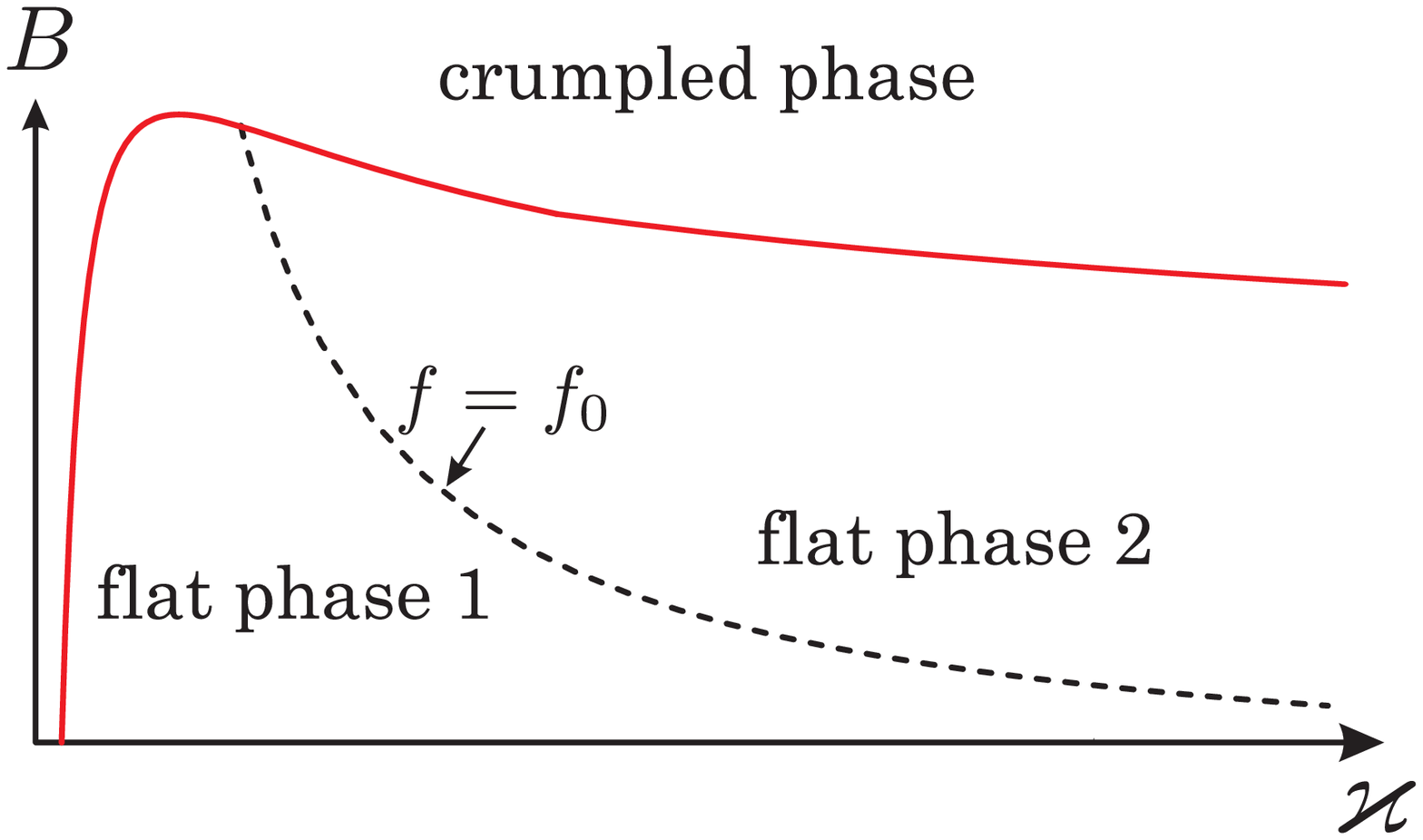} }
\caption{
  Phase diagram of graphene in the plane of parameters $\varkappa$ (bending rigitidy) and $B$ (disorder) at non-zero tension $\sigma$.
   The CT separating crumpled and flat phases is shown by full red line.
     Line $f=f_0$ 
  separates two flat phases with different behavior of dynamical and static correlation functions.  }
\label{FS1}
\end{figure}

In both phases, ${\varkappa_qq^2\ll\sigma}$, so that the scaling of $\varkappa$ is irrelevant
for the CT and the membrane remains flat.
The phases can be distinguished by the behavior of dynamical and static
fluctuations.  To characterize  both types of fluctuations
  we introduce the corresponding
 correlation  functions \cite{disorders-Morse-Grest,my-crump}:
\begin{eqnarray}
&&{H^{\rm d}(x)=\overline{\left\langle\p_\alpha{\mathbf h}(0)\p_\alpha{\mathbf h}(\mathbf x)\right\rangle}-H^{\rm s}(x)},
\\
&&{H^{\rm s}(x)=\overline{\left\langle\p_\alpha{\mathbf h}(0)\right\rangle\left\langle\p_\alpha{\mathbf h(\mathbf x)}\right\rangle}}.
\end{eqnarray}
Here, ${\langle\ldots \rangle}$ stands for averaging over dynamical
fluctuations in a given disorder realization, while the overbar means averaging over disorder.
Functions  $H^{\rm d}_q$ and  $ H^{\rm s}_q$ are given by the first and
second term in Eq.~(14) of the main text, respectively. For $q\ll \tilde{q}_\sigma,$ we get
\begin{eqnarray}
&&H_q^{\rm d}\! \approx\!  \frac{d_c T }{\sigma}\left(1-\frac{\varkappa_q q^2}{\sigma}\right),
\\
&&H_q^{\rm s}\! \approx\! \frac{d_c T f_q \varkappa_q q^2}{\sigma^2}\left(1-\frac{2\varkappa_q q^2}{\sigma}\right).
\end{eqnarray}
Here we keep corrections with respect to the small parameter $\varkappa q^2/\sigma$.
Solving  Eq.~(23) of the main text, we find that the leading contributions to both dynamical and
static functions are regular,
$$
{H_q^{\rm d(0)}=d_cT/\sigma=\text{const}(q)}$$
and
$${H_q^{\rm s (0)}=d_cTf_q\varkappa_q q^2/\sigma^2\propto q^2},$$
each behaving in the same way for ${f>f_{\rm cr}}$ and ${f<f_{\rm cr}}.$
However, the corrections
show an anomalous scaling different in the two phases:
\BEA
\delta H_{q\to0}^{\rm d}&=&-\frac{d_c T}{\sigma^2} \varkappa_q q^2 \propto \left\{  \begin{array}{c}
                             q^{2-2\eta},\,\,\,\, \text{for}\,\,\,\, f<f_{\rm cr} \\
                             q^{2+\eta/6},\,\,\,\,\,\, \text{for}\,\,\,\, f>f_{\rm cr},
                           \end{array}
\right.
\nonumber
\\
\nonumber
\delta H_{q\to0}^{\rm s}&=&-\frac{2d_c T}{\sigma^3}f_q \varkappa_q^2 q^4 \propto \left\{  \begin{array}{c}
                             q^{4-2\eta},\,\,\,\, \text{for}\,\,\,\, f<f_{\rm cr} \\
                             q^{4+\eta/6},\,\,\,\,\,\, \text{for}\,\,\,\, f>f_{\rm cr}.
                           \end{array}
\right.
\\
\label{fcrit}
\EEA
These corrections are obtained by differentiating the correlation functions with respect to $\sigma$:
${\delta H_{q}^{\rm d}=\p(\sigma H_{q}^{\rm d})/\p\sigma},$  ${\delta H_{q}^{\rm s}=\p(\sigma^2 H_{q}^{\rm s})/\p\sigma}. $
The difference in the behavior of the correlation functions (\ref{fcrit}) distinguishes the two flat phases,
flat phase 1 and flat phase 2, shown in Fig.~S1.

\subsection{Comparison with numerical simulations of Ref.~\cite{katsnelson16} }

Here, we present some details on the comparison of our theory with numerical simulations for clean graphene presented in Ref.~\cite{katsnelson16}. The simulations were performed for graphene samples of various sizes.
Phenomenological formulas were found that fitted very well numerical data points in the considered range of stress and strain, see Eqs.~(10), (11) of Ref.~\cite{katsnelson16}.  In our notations, these formulas become
\BEA
k_{\rm eff}&=&\frac{2 \xi [k_0/\xi_0+ C D (\xi-\xi_0)] }{1+D (\xi-\xi_0)}, \label{keff}
\\
\sigma&=&\frac{2}{D} \left(  \frac{k_0}{\xi_0} -C\right) \ln[1+D(\xi-\xi_0) ]
\nonumber
\\
& +&2 C(\xi-\xi_0) , \label{sigma0}
\EEA
The parameters $C,~ D,~ k_0,~\xi_0$ were found to be dependent on the system size $L$; this dependence was fitted by phenomenological formulas  presented in Table I  of Ref.~\cite{katsnelson16}.   For comparison with the theory, we used  the values of these parameters for the sample of largest size (37888 atoms) considered in Ref.~\cite{katsnelson16}, since these numerical data should be less affected by finite-size effects.

 Phenomenological equations \eqref{keff} and \eqref{sigma0} yield the implicit dependence of  effective stiffness $k_{\rm eff}$ on $\sigma$, as found numerically in Ref.~\cite{katsnelson16}.  It is worth noting that $k_{\rm eff}$ does not go to zero at the point $\xi=\xi_0$   of zero strain (and zero stress).
In other words,  numerical simulations show the existence of small  but finite  linear stiffness $k_{\rm eff}(0)$. This is a finite-size effect: the power-law renormalization of effective stiffness is cut off by the system size. Indeed, the data of Ref.~\cite{katsnelson16} also shows that this residual stiffness vanishes in the limit of large systems, $L\to\infty$. To compare the numerical data with our theoretical prediction for $k_{\rm eff} (\sigma)$ given by Eq.~(19) of the main text (that corresponds to the thermodynamic limit), we have removed this finite-size effect by shifting Eq.~\eqref{sigma0} in such a way that the large-$L$ condition $k_{\rm eff}(0)=0$ is restored.



\begin{thebibliography}{99}
\bibitem{comm} In these papers, HL was written in terms of tension $\sigma'=L^{-2} \p  F /\p \xi= 2\xi \sigma.$
\end{thebibliography}
 \end{document}